\documentclass[
    ,final            
  ]
  {aipproc}

\layoutstyle{6x9}
\newcommand{\grado}{^{\circ}}
\usepackage{epsfig}

\begin{document}

\title{The search for DM in nearby dSph galaxies with MAGIC: candidates, results and prospects\footnote{Talk presented by M.A.S.C. at the 8th UCLA Dark Matter Symposium, Marina del Rey, USA, 20-22 February 2008}}


\author{Miguel A. S\'anchez-Conde}{
  address={Instituto de Astrof\'isica de Andaluc\'ia (CSIC), E-18008 Granada, Spain}
}
\author{A. Biland}{
  address={Institute for Particle Physics, ETH Zurich, CH-8093 Zurich, Switzerland}
}
\author{M. Doro}{
  address={Universita di Padova and INFN, I-35131 Padova, Italy}
}
\author{S. Lombardi}{
  address={Universita di Padova and INFN, I-35131 Padova, Italy}
}
\author{D. Nieto}{
  address={Universidad Complutense, E-28040 Madrid, Spain}
}
\author{F. Prada}{
  address={Instituto de Astrof\'isica de Andaluc\'ia (CSIC), E-18008 Granada, Spain}
}
\author{M. Rissi}{
  address={Institute for Particle Physics, ETH Zurich, CH-8093 Zurich, Switzerland}
}
\author{L. S. Stark}{
  address={Institute for Particle Physics, ETH Zurich, CH-8093 Zurich, Switzerland}
}
\author{F. Zandanel$^*$ for the MAGIC collaboration}{
  address={http://wwwmagic.mppmu.mpg.de}
}

\begin{abstract}
At present, dwarf spheroidal galaxies satellites of the Milky Way may represent the best astrophysical objects for dark matter (DM) searches with gamma-ray telescopes. They present the highest mass-to-light ratios known in the Universe. Furthermore, many of them are near enough from the Earth to be able to yield high predicted DM annihilation fluxes that might be observed by current gamma-ray instruments like MAGIC. The picture has become even better with the recent discovery of new dwarfs. These new objects are expected to yield even higher DM annihilation fluxes, since most of them are nearer than the previously known dwarfs and are even more DM dominated systems. Here a tentative list of the best candidates is given. The observational results obtained with MAGIC from the Draco dwarf as well as the observation of other dwarfs carried out by other Cherenkov telescopes are presented as well. Finally, we discuss the detection prospects of such kind of objects in the context of DM searches.
\end{abstract}

\maketitle


\section{Introduction}

In the Lambda Cold Dark Matter ($\Lambda$CDM) paradigm, which represents the most accepted cosmological scenario at present, around 23\% of the Universe consists of non-barionic Dark Matter (DM) \citep{wmap}. This kind of DM is needed in order to explain a large amount of astrophysical processes at all scales, such as the formation of structures in the early Universe, their evolution with time so that they can form the structures observable today, the results from gravitational lensing, the rotation curves of individual galaxies, etc. \citep{review} To be able to explain these astrophysical puzzles, the DM particles should interact very weakly with the ordinary matter and should have low thermal velocities. A very good candidate that fulfill the above mentioned characteristics arises in the context of SUSY (SuperSymmetry, the most preferred Particle Physics scenario beyond the Standard Model). Effectively, in R-parity conserving supersymmetric theories, the lightest SUSY particle (LSP) remains stable, and the widely studied minimal supersymmetric extension of the standard model of particle physics (MSSM) can predict a neutralino as the LSP with a relic density compatible with the WMAP bounds (see e.g.~Ref.~\citep{review2}). An interesting property of the neutralino is that it is predicted to be its own antiparticle, which means that it will annihilate when interacting with other neutralinos. This fact is crucial for detectability purposes and represents indeed the main vehicle used at present in DM searches. One of the products of these annihilations are predicted to be gamma-rays, whose specific energy will vary according to the chosen particle physics model, but that is expected to lie in the energy range covered by the current Imaging Atmospheric Cherenkov Telescopes (IACTs) like CANGAROO, HESS, MAGIC and VERITAS, and/or by the Fermi (formerly GLAST) satellite, sucessfully launched in June 11th, 2008. The DM annihilation flux is proportional to the square DM density, which means that the best places to search for DM in the Universe are those with the highest expected DM densities. Distance is also very important, since high DM dominated systems that are located too far from us will yield too low DM annihilation fluxes at Earth. Having both considerations in mind, dwarf spheroidal (dSph) galaxies arise as very good candidates for DM searches, as we will discuss later in detail.

\section{The MAGIC telescope}

The MAGIC telescope is a single-dish 17~m IACT located at the Roque de los Muchachos Observatory in the Canary Islands (28.8 N, 17.8 W, 2200m a.s.l.). In the standard trigger operation mode, it reaches an energy threshold as good as $\sim$55 GeV for small zenith angles, which represents the lowest energy threshold over all the current IACTs. This fact, together with its very good sensitivity of 1.6\% of the Crab in 50h, makes MAGIC a very competitive instrument for DM searches. MAGIC has an energy resolution around 20-30\% \citep{energyresol} and an angular resolution $\sim$0.1$\grado$. The camera, composed of 577 photomultiplier tubes, gives a FoV of 3.5$\grado$. It is worth mentioning that very recently a new trigger system (the so-called SUM-trigger) has been developed which allows the detection of low energy gamma rays above 25 GeV \citep{sumtrigger}. Furthermore, a second MAGIC telescope of equal dimensions is expected to enter in operation in 2009. With both telescopes working in steoroscopic mode, the gamma/hadron separation will improve significantly. This should allow to reach a better sensitivity and even a lower energy threshold.

\section{The DM gamma signal} \label{sec:DMsignal}

Let us make the assumption that the neutralino is the main component of the DM present in the Universe. Then, the expected total number of continuum $\gamma$-ray photons received per unit time and per unit area, above the energy threshold E$_{th}$ of the telescope, observing at a given direction $\Psi_0$ relative to the centre of the DM halo is given by:

\begin{equation}
F(E>E_{\rm th})=\frac{1}{4\pi} {f_{SUSY}} \cdot U(\Psi_0) 
\label{eq1}
\end{equation}

The factor $f_{SUSY}$ encloses all the particle physics. At leading order, neutralinos ($\chi$) do not annihilate into two-body final states containing photons. However, at one loop it is possible to get processes such as $\chi+\chi \rightarrow \gamma \gamma \nonumber$, $\chi+\chi \rightarrow Z \gamma$ with monochromatic outgoing photons of energies $E_\gamma\sim m_\chi,\;\,\ E_\gamma\sim m_\chi- {m_Z^2\over{4 m_\chi}}$ respectively \cite{Bergstrom:1997fh}. Also, the neutralino annihilation can produce a continuum $\gamma$-ray spectrum from hadronization and subsequent pion decay which can dominate over the monochromatic gammas. Taking all of this into account, $f_{SUSY}$ is calculated as:

\begin{equation}
f_{SUSY}=\frac{\theta(E_{th}>m_{\chi})\cdot 2 \left<v \sigma_{\gamma
\gamma}\right>+\theta(E_{th}>m_{\chi}-{m_Z^2\over{4m_\chi}})\cdot
\left<v \sigma_{\gamma Z}\right>+~k\left<v \sigma_{cont.}\right>}{2 m_{\chi}^2}, 
\end{equation} 

\noindent where $<v \sigma_{\gamma\gamma}>$, $<v \sigma_{\gamma Z}>$ and $<v \sigma_{cont.}>$ are the annihilation cross sections of each of the above mentioned processes, $\theta$ is the step function and $k$ the photon
multiplicity for each neutralino annihilation. 

All the astrophysical considerations are included in the expression $U(\Psi_0)$ in Eq.(\ref{eq1}). This factor accounts for the DM distribution, the geometry of the problem and also the beam smearing of the IACT, i.e. $U(\Psi_0)=\int J(\Psi)B(\Omega)d\Omega$, where $B(\Omega)d\Omega$ represents the beam smearing of the telescope, commonly known as the Point Spread Function (PSF), and $J(\Psi)$ represents the integral of the line-of-sight of the square of the DM density along the direction of observation $\Psi$. In Ref.~\citep{masc} a more detailed explanation of each of these terms can be found.

\section{Search of DM in dwarf galaxies}

As already commented, dSph galaxies satellites of the Milky Way are very good candidates for DM searches: 
\begin{enumerate}
 \item DSphs are the most DM dominated systems known in the Universe. Their inferred mass-to-light (M/L) ratios can be as high as 1000 for some of the recently discovered dSphs, according to detailed studies of the kinematics of their member stars. The high density in dSph galaxies reflects the high density of the Universe at the time of their formation. They also conform to the general (anti)correlation of DM content and central halo density with luminosity observed in disk galaxies \citep{persic}.
 \item They are relatively near from us. Indeed, many of them are located nearer than 100 kpc (e.g. Draco, UMi and most of the new SDSS dwarfs). This fact becomes crucial in order to obtain a high DM annihilation flux at Earth.
 \item Most of them are expected to be free from bright astrophysical gamma sources. This means that any gamma signal that we might receive in our telescopes coming from these dwarfs will potentially be due to an effective DM annihilation. This is in contrast with what it is expected to occur for other DM candidates, like the Galactic Center, nearby galaxies or very massive galaxy clusters.
\end{enumerate}

Nowadays, we know the existence of at least 24 Milky Way satellites, more than half of them discovered in the last few years using SDSS data \citep{strigarinature}. In Table \ref{tab:dwarfs} we show a tentative list of those dSph galaxies that could be the best candidates for DM searches, according to its distance and/or M/L ratio. Note that the recently discovered SDSS dwarf galaxies listed here (Coma, UMa~II, Willman~1 and Segue~1) have typically higher M/L ratios and are closer than the previously well-known dwarfs. Both facts should increase the chance of detection. 

\begin{table}
\begin{tabular}{lccclr}
\hline
  \tablehead{1}{c}{b}{DSph}
  & \tablehead{1}{c}{b}{D$_{\odot}$ (kpc)}
  & \tablehead{1}{c}{b}{L ($10^{3}~L_{\odot}$)}
  & \tablehead{1}{c}{b}{M/L ratio}
  & \tablehead{1}{c}{b}{Reference}
  & \tablehead{1}{c}{b}{Best positioned \\IACTs}   \\
\hline
Carina & 101 & 430 & 40 & \citep{mateo98} & HESS, CANGAROO\\
Draco & 82 & 260 & 320 & \citep{mateo98}  & {\bf MAGIC}, VERITAS\\
Fornax & 138 & 15500 & 10 & \citep{mateo98}  & HESS, CANGAROO\\
Sculptor & 79 & 2200 & 7 & \citep{mateo98}  & HESS, CANGAROO\\
Sextans & 86 & 500 & 90 & \citep{mateo98}  & HESS, CANGAROO\\
UMi & 66 & 290 & 580 & \citep{mateo98} & {\bf MAGIC}, VERITAS\\
Sagittarius\tablenote{Not a dSph, but listed here because of its traditional interest for DM searches.} & 24 & 58000 & 25 & \citep{mateo98,helmiwhite01}  & HESS, CANGAROO\\
Coma Berenices & 44 & 2.6 & 450 & \citep{simongeha07} & {\bf MAGIC}, VERITAS\\
UMa II & 32 & 2.8 & 1100? & \citep{simongeha07} & {\bf MAGIC}, VERITAS\\
Willman 1 & 38 & 0.9 & 700 & \citep{simongeha07} & {\bf MAGIC}, VERITAS\\
Segue 1\tablenote{This dwarf did not appear in the talk, since it was not confirmed at that moment.} & 23 & 0.3 & >1320 & \citep{geha08} & {\bf MAGIC}, VERITAS\\
\hline
\end{tabular}
\caption{A list of dSph satellites of the Milky Way that may represent the best candidates for DM searches according to their distance and/or inferred M/L ratio.}
\label{tab:dwarfs}
\end{table}

\subsection{Observations of Draco with MAGIC}

The Draco galaxy, a satellite of the Milky Way, represents one of the best candidates to search for DM outside our galaxy \citep{evans}. It is near (82 kpc) and it has probably more observational constraints than any other dSph galaxy. The latter becomes crucial when we want to make realistic predictions of the expected observed $\gamma$-ray flux due to DM annihilation. Furthermore, its inferred M/L ratio $\sim$300 indicates that it is highly DM dominated, and its position in the Northern Sky is excellent to carry out high quality observations with the MAGIC telescope.

\begin{figure}
   \centering
   \begin{minipage}[b]{0.49\textwidth}
   \centering
	\includegraphics[height=5.8cm,width=7cm]{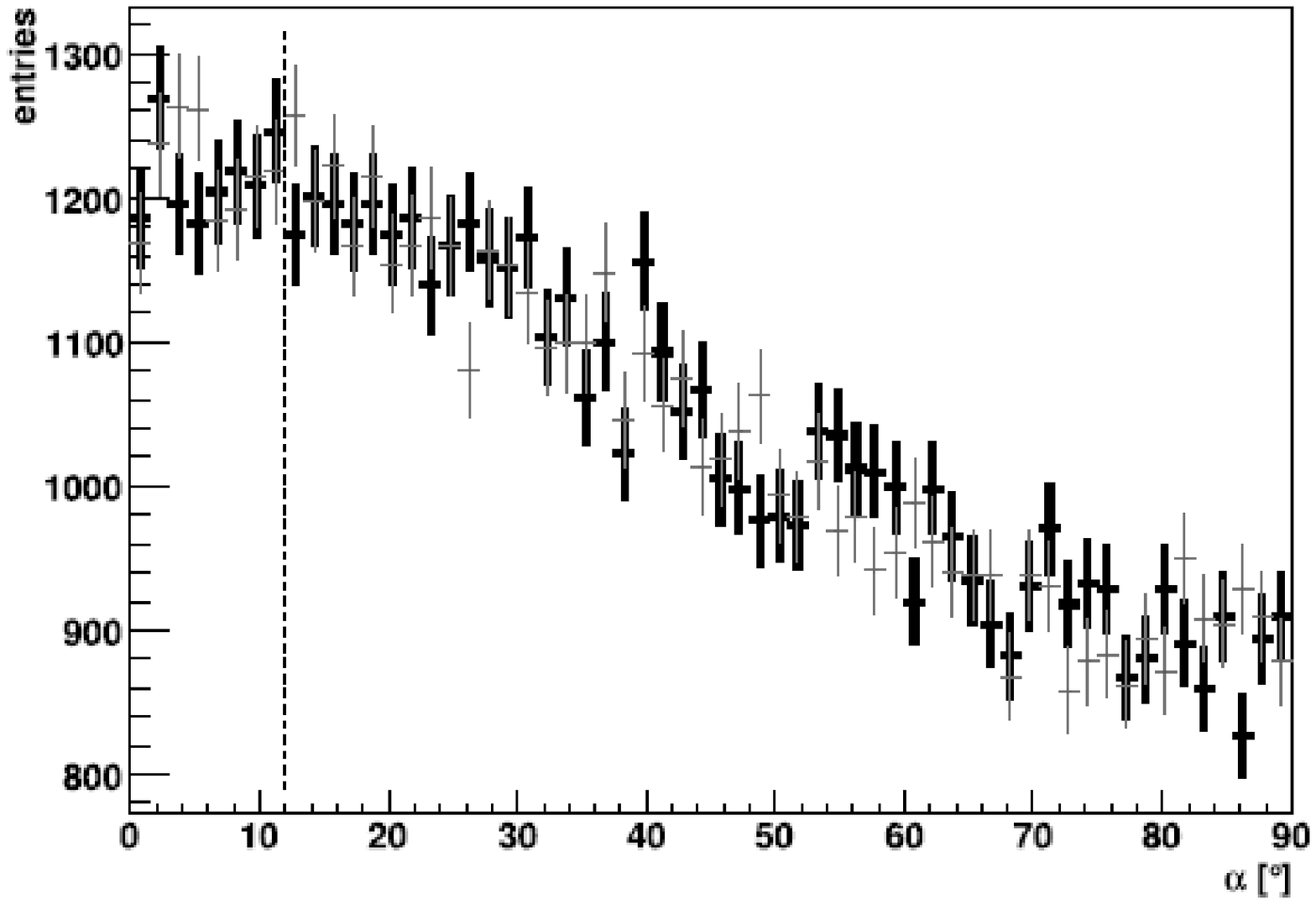}
  \end{minipage}
  \begin{minipage}[b]{0.49\textwidth}
  \centering
	\includegraphics[height=5.8cm,width=7cm]{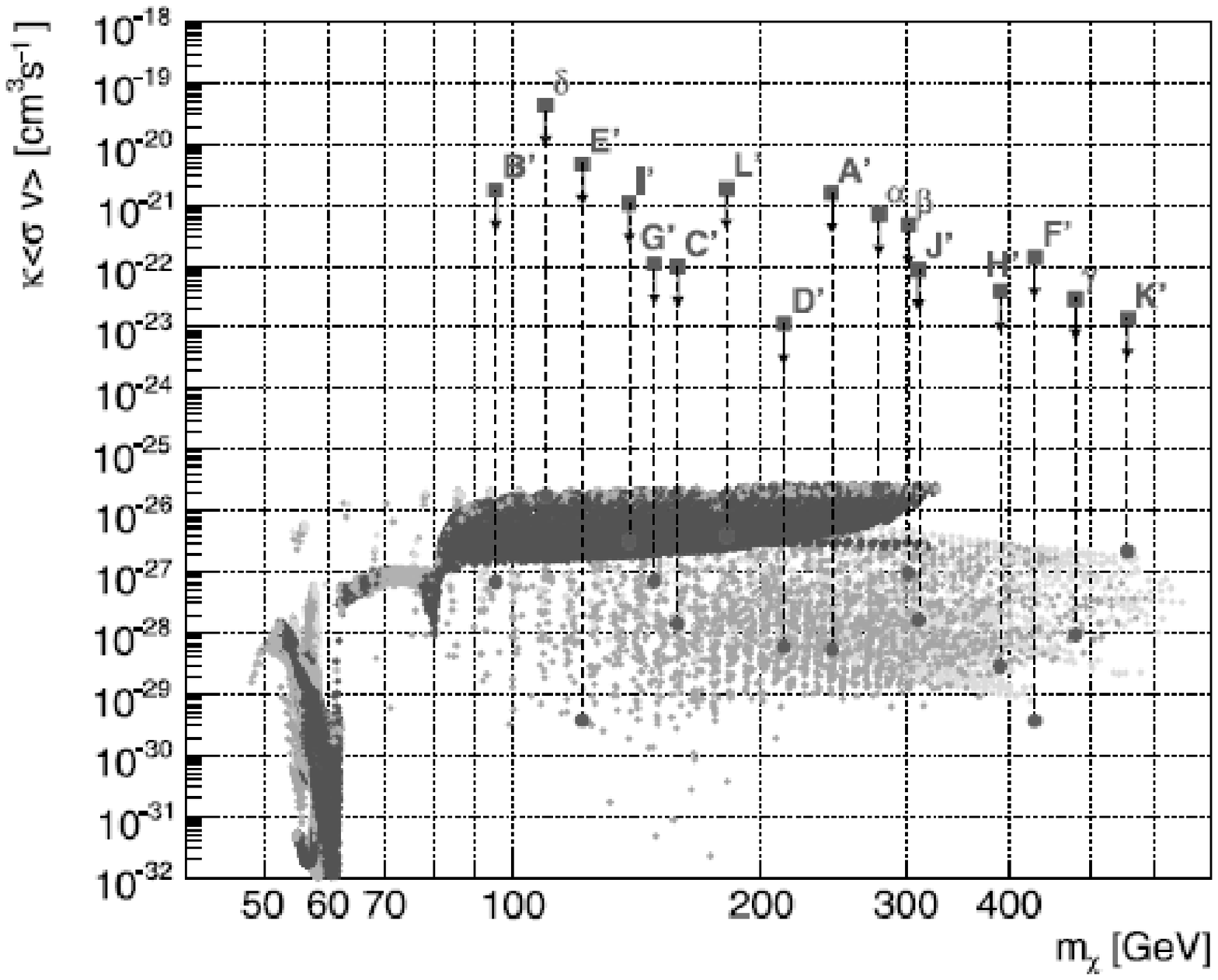}
	\caption{Left panel: Distribution of the $\alpha$ parameter for $\gamma$-ray candidates coming from the center of Draco (bold face marker) and background (light markers) for data taken between 2007 May 9 and May 20. The energy threshold is 140 GeV. The signal region was supposed to be that with $\alpha$ < 12$\grado$. Right panel: Thermally averaged neutralino annihilation cross section as a function of the neutralino mass for mSUGRA models. Dots with roman letters represent benchmark models in Ref.~\citep{battaglia}, while dots with greek letters are models chosen in Ref.~\citep{MAGICdraco}. The squares indicate the flux upper limit in units of $<\sigma~v>$, assuming the parameters given in Ref.~\citep{masc} for the DM density profile.}
	\end{minipage}
  \label{fig:magic_obs} 
\end{figure}

MAGIC observed this dwarf for a total observation time of 7.8 hours during May 2007 \citep{MAGICdraco}. We performed a standard analysis of the data. The standard analysis procedure, described in detail e.g.~in Ref.~\citep{analisis}, essentially consists on the parameterization of each event using the Hillas image parameters \citep{hillas} and on a posterior hadronic background supression and energy estimation by means of the Random Forest method. A reference data sample of a very well-known object is used for optimizing the gamma/hadron cuts (typically the Crab Nebula, which represents the ``standard candle'' in $\gamma$-ray astronomy). We show in the left panel of Fig.~\ref{fig:magic_obs} the distribution of the ALPHA Hillas parameter (see e.g. Refs.~\citep{analisis,hillas} for a detailed description of this parameter): no gamma signal was found above an Energy threshold of 140 GeV. We derived an upper limit for the flux (2$\sigma$ level) to be around 1.1 x 10$^{-11}$~ph~cm$^{-2}$~s$^{-1}$, assuming a power-law with spectral index -1.5 and a point-like source.

To estimate how far we are from a successful DM signal detection, first we will need to make some theoretical predictions of the expected DM annihilation flux. We assumed a DM distribution model for Draco well described by the formula $\rho_{\rm d}(r) = C r^{-\alpha} {\rm exp} (-\frac{r}{r_{\rm b}})$ proposed by \citep{kmmds}. We considered two cases, the profile with a cusp $\alpha=1$ and a core $\alpha=0$, with those parameters given in Ref.~\citep{masc}. To include the particle physics in the calculations, we chose the benchmark models proposed in Ref.~\citep{battaglia}. Doing so, we obtained that our upper limits from MAGIC observations are $\sim$10$^3$-10$^9$ above the predicted values (see right panel of Fig.\ref{fig:magic_obs}). Although still very far, our results can exclude a very high DM annihilation flux. We note here that first-order radiative corrections \citep{Bringmann08} and/or the role of substructure \citep{kuhlen08}, which may boost the signal, were not yet included in the calculations.

\subsection{Results from other experiments}

In addition to MAGIC, there are also other IACTs that have already carried out observations of some of the Milky Way satellite galaxies in the context of DM searches. This is the case of H.E.S.S., which observed Saggitarius for 11 hours exposure time in June 2006 \citep{SagHESS}. No significant excess was found above E$_{th}>$250 GeV, although they claimed the exclusion of some pMSSM models from the data assuming a core DM density profile for this galaxy. The Whipple IACT also observed two galaxy satellites during 2003: Draco (14.3 h) and UMi (17.2 h). No significant excess was found above 400 GeV, and it was not possible to rule out any of the MSSM allowed models \citep{whippleobs}. 

\section{A MAGIC future}

MAGIC will continue searching for DM signals in dSph galaxies\footnote{At the moment of this talk, MAGIC had only observed Draco in the context of DM searches, but some time later (Spring 2008) also the Willman~1 dSph was observed for a total of 15.5 hours. Again, no signal excess was found, although the derived upper limits for the flux were more stringent in this case \citep{MAGICwillman1}.}. Furthermore, MAGIC II will improve our chances of detection and will impose more stringent limits to the particle physics models (Draco and Willman~1 detection prospects for MAGIC~II can be found in Ref.~\citep{doro}). However, it is important to obtain more observational data in order to build more accurate DM density profiles. The possible role of substructure as a boost factor of the flux has to be clarified as well. However, it seems that other dSph galaxies will not change the expected DM annihilation flux drastically (see e.g.~\citep{Pieri08}). From the theoretical side, it is expected that the LHC at CERN will reduce the uncertainties coming from particle physics. Furthermore, commonly neglected first-order radiative corrections may be very important, so also the theoretical SUSY predictions are probably not yet final.


\end{document}